\documentclass[preprint,showpacs,preprintnumbers,amsmath,amssymb]{revtex4}

\usepackage{graphicx}
\usepackage{dcolumn}
\usepackage{bm}

\begin{document}

\preprint{APS/123-QED}

\title{The pairing state in KFe$_2$As$_2$ studied by measurements of the magnetic vortex lattice}

\author{H. Kawano-Furukawa}
\affiliation{Division of Natural/Applied Science, G.S.H.S., Ochanomizu University, Bunkyo-ku, Tokyo 112-8610,
Japan}
\author{C. J. Bowell}
\affiliation{Dept. of Materials Science and Metallurgy, Univ. of Cambridge, Pembroke Street, Cambridge CB2 3QZ,
United Kingdom}
\author{J. S. White}
\affiliation{Laboratory for Neutron Scattering, ETH Zurich and Paul Scherrer Institut, CH 5232 Villigen PSI,
Switzerland}
\author{R.W. Heslop}
\author{A.S. Cameron}
\author{E.M. Forgan}
\affiliation{School of Physics and Astronomy, Univ. of Birmingham, Edgbaston, Birmingham B15 2TT, United
Kingdom}
\author{K. Kihou}
\author{C. H. Lee}
\author{ A. Iyo}
\author{H. Eisaki}
\affiliation{National Institute of Advanced Industrial Science and Technology (AIST), Tsukuba, Ibaraki 305-8568,
Japan}
\affiliation{JST, Transformative Research-Project on Iron Pnictides (TRIP), Chiyoda, Tokyo 102-0075, Japan}
\author{T. Saito}
\affiliation{Department of Physics, Chiba University, Chiba 263-8522, Japan}
\author{H. Fukazawa}
\author{Y. Kohori}
\affiliation{JST, Transformative Research-Project on Iron Pnictides (TRIP), Chiyoda, Tokyo 102-0075, Japan}
\affiliation{Department of Physics, Chiba University, Chiba 263-8522, Japan}
\author{R. Cubitt}
\author{C. D. Dewhurst}
\affiliation{Institut Laue-Langevin, 6 rue Jules Horowitz, 38042 Grenoble, France}
\author{J. L. Gavilano}
\affiliation{Laboratory for Neutron Scattering, ETH Zurich and Paul Scherrer Institut, CH 5232 Villigen PSI,
Switzerland}
\author{M. Zolliker}
\affiliation{Laboratory for Developments and Methods, Paul Scherrer Institute, CH-5232 Villigen, PSI, Switzerland}

\date{tentative version \today}

\begin{abstract}

 We report the observation, by small-angle-neutron-scattering (SANS), of magnetic flux lines - ``vortices'' in super-clean KFe$_2$As$_2$ single crystals. The results show clear Bragg spots from a well ordered vortex lattice, for the first time in a FeAs superconductor. These measurements can give important information about the pairing state in this material, because the spatial variation of magnetic field in the vortex lattice reflects this pairing.  With field parallel to the fourfold $c$-axis, nearly isotropic hexagonal packing of vortices was observed without VL-symmetry transitions up to high fields, indicating rather small anisotropy of the superconducting properties around this axis. This rules out gap nodes parallel to the $c$-axis, and thus $d$-wave and also anisotropic $s$-wave pairing. The strong temperature-dependence of the scattered intensity down to $T~\ll~T_c$ further indicates either widely different full gaps on different Fermi surface sheets, or nodal lines perpendicular to the axis.

\end{abstract}

\pacs{74.25.Uv, 74.70.Xa, 74.20.Rp, 74.25.-q}
\maketitle

Understanding the mechanism and symmetry of electron pairing in iron-based superconductors represents an
important challenge in condensed matter physics. Soon after the discovery of high temperature superconductivity in iron pnictide materials, many theorists
suggested that pairing in these materials arises from a magnetic interaction~\cite{R1}. It is proposed that in
optimally-doped materials, there is a repulsive interaction between holes on Fermi surface (FS) sheets at the
center of the Brillouin zone and the almost equal number of electrons on other sheets at the corners. The simplest
form of pairing that could arise from this interaction has the opposite sign of the pairing order parameter for the
two types of carrier. Such a pairing state may be described as extended $s$-wave or $s_{\pm}$ since the order
parameter has the {\it same} symmetry as the crystal, and is nonzero on all parts of the FS, despite changing sign. In
this respect it may be contrasted with $d$-wave pairing in cuprate materials, in which the order parameter has a
lower symmetry than the underlying crystal, changes sign around a single sheet of FS and therefore has nodes.
Other pairing states have been proposed for the pnictides~\cite{R2, R3} which maintain s$_{\pm}$ symmetry, but
in which the order parameter also changes sign on at least one FS sheet. Our material,  KFe$_2$As$_2$,
belongs to the `122' family of iron pnictide superconductors; the optimally-doped BaFe$_2$(As$_{1-x}$P$_x$)$_2$ member shows
clear indications of nodes in the order parameter~\cite{R4, R5} which may imply either nodal $s_{\pm}$ pairing or
alternatively $d$-wave gap symmetry~\cite{R3}. On the other hand, in optimally-doped Ba$_{1-x}$K$_x$Fe$_2$As$_2$ with $T_c$ $>$ 35 K, the order parameter appears to be fully-gapped on all parts of the FS, although with
large and small gaps on different FS sheets~\cite{R6, R7}. Optimally-doped BaFe$_{2-x}$Co$_x$As$_2$ is similarly fully-gapped~\cite{R7A}. Such observations raise the question whether all the iron-based superconductors, even within the 122 family, have the same gap symmetry and pairing mechanism. We note, however that nodal gaps may be observed in the `clean' materials with long mean free paths (which allow dHvA observations~\cite{R7B,R11}), and full gaps may be observed in `dirtier' materials~\cite{R6,R7,R7A}. This indicates that low-energy states due to electron scattering are not a major cause of `nodal' behavior.

The vortex state in superconductors reflects the electron pairing state. However, SANS experiments up till now on
pnictide superconductors have shown disordered VLs~\cite{R8, R9, R10}. It is likely that nano-scale compositional
inhomogeneity is a major source of vortex pinning and disorder in doped crystals.
In KFe$_2$As$_2$, with Ba fully replaced by K, we have an over-doped but clean 122 material. 
Here we show how SANS measurements on the vortex lattice (VL) in this material can make a contribution to the debate on the gap structure in pnictide superconductors.

For the present study, single crystals of stoichiometric KFe$_2$As$_2$ were grown by a flux method. Detailed methods are described elsewhere~\cite{RS}. SQUID magnetization measurements reveal that $T_c$ is 3.6 K with a 10 - 90$\%$ width of 0.2 K, and resistivity measurements give a typical residual resistance ratio $\sim$ 400~\cite{RS}. These facts indicate the system is very clean and indeed dHvA signals are observed~\cite{R11}.

SANS experiments were performed on instruments D11 and D22 at the ILL France and
SANS-I at the SINQ facility in PSI, Switzerland. A mosaic of co-aligned crystals with the $c$-axis and one of the
tetragonal [100] axes in a horizontal plane was glued to aluminium sample-plates with hydrogen-free CYTOP$^\circledR$
varnish and cooled to a minimum of 1.5 K using a $^4$He cryostat at D11 and to 50 mK using dilution
refrigerators at D22 and SANS-I. The total mass of the crystals for the ILL experiments was about 100 mg but 250
mg at PSI. Using a cryomagnet, fields from 0.1 to 0.9 T were applied parallel to the crystal $c$-axis and
approximately parallel to the incident neutron beam. Neutrons with a wavelength of $\lambda_n =$ 10 $\rm{\AA}$ were
used and a position-sensitive detector was set at 15 m (ILL) and 18 m (PSI) from the sample position to obtain a
reasonable $q$ resolution. Except when measuring the main beam intensity, the undiffracted beam was caught on
a neutron-absorbent beamstop.
The sample had sufficiently weak pinning that the VL could be formed by applying the field well below $T_c$;
field-cooling through $T_c$ was not required. Indeed, the VL perfection was slightly improved by oscillating the
value of the field by $\sim$ 0.01 T about its final value at base temperature, before taking measurements. To
observe the VL diffraction, the sample and cryomagnet were rocked together either vertically or horizontally by
small angles about the $B$ $\parallel$ beam direction to bring the various diffraction spots through the Bragg condition.
Backgrounds were taken with no VL present, either by heating above $T_c$ or by removing the applied field. Note
that the real-space VL nearest-neighbor positions may be visualized by rotating the diffraction patterns by $90^{\circ}$ about the field axis and adding another spot at the central main beam position.
The SANS data were displayed and rocking curves were analyzed using the GRASP analysis package~\cite{RDew1}.

Figures 1a-d show VL diffraction patterns with clear Bragg spots, measured at $T =$ 50 mK and 1.5 K at selected
magnetic fields applied parallel to the fourfold [0 0 1] crystal axis and approximately parallel to the neutron beam.
The 12 spots arise from a mixture of two domain orientations of hexagonal VLs, which have nearest-neighbor
vortices along the tetragonal [1 0 0] and [0 1 0] axes, and are degenerate~\cite{RS}.
Although the angles of the spots change with increasing field (Fig. 1e) and the observed intensity becomes
weaker, the VL symmetry does not change for all conditions under which a VL signal is observed (see Fig. 1f)~\cite{FN}. In
this respect, KFe$_2$As$_2$ is unlike almost all conventional and unconventional superconductors with a
fourfold axis~\cite{R12, R13, R14, R15, R16}. In conventional materials, anisotropy of the Fermi velocity leads to
VL phase transitions as the field is increased~\cite{R17}, so we conclude that the FS anisotropy is weak in KFe$_2$As$_2$. First-principles calculations using Eilenberger theory show that gap anisotropy has similar effects~\cite{R18, R19}, and that
a $d$-wave nodal gap gives rise to a square VL at a small fraction ($\sim$ 0.15 at $T = 0.5T_c$) of $B_{c2}$~
\cite{R20}. These calculations are confirmed by experiments on $d$-wave materials~\cite{R15, R16}. Since in KFe$_2$As$_2$, a square VL is absent up to 0.9 T - approximately half the upper critical field $B_{c2}$ at 50 mK, we
rule out a nodal order parameter with the gap varying around the basal plane.

Figure 2a shows the $B$-field-dependence of the VL `Form Factor' $(F)$ derived from the integrated intensity~\cite{RS} of the `on axis' VL diffraction spots at 50 mK and at 1.5 K. $F$ is a measure of the
spatial variation of the field inside the mixed state; it generally decreases at large field as the vortex cores begin to overlap. These data were fitted to a modified London model with core/non-local corrections~\cite{R21}:
\begin{equation}
  F = \frac{B}{1+q^2\lambda^2} \exp(-cq^2\xi^2)
\end{equation}

Here, $\lambda$ is the London penetration depth, $\xi$ is the coherence length and $c$ is an empirical core cut-
off parameter. The good fit suggests a conventional field-dependence without multi-gap effects~\cite{R22} and
allows us to extrapolate to obtain the zero-field value of $\lambda$. The best fit at 50 mK gave $\lambda$ = 203
nm and $c$ = 0.52, using $\xi$ = 13.5 nm (from $B_{c2}$ = 1.8 T), and at 1.5 K: $\lambda$ = 240 nm and $c$ =
0.55, with $\xi$ = 15.9 nm (from $B_{c2}$ = 1.3 T). Random errors in $\lambda$ were small compared with probable
systematic (calibration) errors $\sim$ 5$\%$.

The low temperature value of $\lambda$ may be combined with normal state properties of KFe$_2$As$_2$~\cite{RS} to confirm that this material has a strongly enhanced carrier mass and that our
samples are `clean', i.e. the electron mean free path, $\ell \gg$ $\xi_0$ , the coherence length.
Furthermore, in Figure 2b, $F$ for both on- and off-axis spots is plotted, but against $q$, rather than field because the VL
distortion gives the off-axis spots a larger $q$ than the on-axis spots. We see that all results lie on nearly the
same line, confirming the essential basal-plane isotropy of the pairing.

Measurements of the intensity of the flux lattice diffraction signal at low fields are equivalent to measurements of the magnetic penetration depth and give its value and temperature-dependence. Hence, the temperature-dependence of the penetration depth (or equivalently superfluid density) is a direct measure of the effects of thermal excitation of quasiparticles over the gap.
Figure 3a shows the temperature dependence of the peak intensity at our lowest field of $B =$ 0.1 T.
Figures 3b-d show fits of these data to different gap functions, (which are described in full in ref.~\onlinecite{RS}). From
Fig. 3b, it is clear that a single full gap cannot fit the low temperature behavior. Also in Fig 3b, we find that a
simple nodal gap, ignoring nonlocal effects~\cite{R23} deviates from the data; however, a single nodal gap
including these effects (Fig. 3c) can reproduce the data very well, with only one additional fitting parameter.
Alternatively, we can obtain a good fit if at least three full gaps of very different magnitudes are included, as shown
in Fig. 3d. Although this uses a rather large number of parameters for the fit, we can not rule out this possibility because KFe$_2$As$_2$ has multiple FS sheets~\cite{R11, R24,R25}, which could support multi-full-gap states. In any case,
the low-temperature behavior is dominated by the minimum values that the gap has at low temperatures; the strong temperature-dependence of the intensity of our diffraction signal, continuing down to at least $T \sim 0.02 T_c$, indicates that KFe$_2$As$_2$ has a range
of gaps extending down to very low values. Probably it has a nodal gap, which is also suggested by recent
penetration depth~\cite{R25} and other measurements~\cite{R26, R27} performed down to various fractions of
$T_c$.

We now consider the fitted values of the ratio $\Delta (T = 0) / k_{B} T_{c}$. The ``largest'' gap in a material may
give a larger ratio than the weak-coupling isotropic BCS value of 1.76: either because of gap anisotropy or strong-
coupling. However, it is not possible for the largest gap in a superconductor to give a smaller ratio. The small values of the maximum full gaps in Figs. 3b \& d argue against gaps without nodes. Even the nodal gap models give values of the ratio close to the weak-coupling limit. We conclude that unlike the optimally-doped compositions, KFe$_2$As$_2$ is a fairly weakly-coupled superconductor.

Finally we discuss the gap structure, bearing in mind that the temperature-dependence of the signal argues for a
nodal state, but the VL structure measurements rule out vertical line nodes in the gap. Note that $T_c$ in the (Ba/
K)Fe$_2$As$_2$ system appears to decrease continuously and monotonically as the doping is increased above
optimal~\cite{R28}. This strongly suggests that the symmetry of the order parameter does not change with doping,
even though the prominent electron sheets in optimally-doped materials are replaced by small hole pockets in
KFe$_2$As$_2$~\cite{R11, R24, R25}. The multi-full-gap scenario may be consistent with this if essentially
isotropic gaps with widely different values develop on each of the FS sheets. Alternatively, we propose an axially
isotropic nodal pairing state to account for our results. In this model, the gap nodes circulate around the
approximately cylindrical sheets of the FS, with the order parameter changing sign as a function of $k_z$, but not
in the $k_x$-$k_y$ plane. For such a state, the pairing interaction should be operative in the $z$-direction, which
requires some three-dimensionality in the electronic structure; dHvA measurements~\cite{R11} and the moderate
anisotropy of $H_{c2}$~\cite{R29} shows that some of the FS sheets in KFe$_2$As$_2$ indeed have a three-dimensional character. We further propose that as the doping in the (Ba/K)Fe$_2$As$_2$ system is increased
beyond optimal, the in-plane pairing interaction becomes weaker, and the nodeless gap changes into a $k_z$-dependent nodal gap. These two states have the same $s_{\pm}$ symmetry, and the node positions are
``accidental'' (not symmetry-determined)~\cite{R30} so the system may go continuously from one to the other as
doping or carrier scattering are varied.

We acknowledge informative discussions with A. Carrington, I. Mazin, K. Machida and A. Maisuradze. Funding:
HK-F is supported by (a) Yamada Science Foundation and (b) Grant-in-Aid for Scientific Research on Innovative
Areas `Heavy Electrons' (No. 20102006) of the Ministry of Education, Culture, Sports, Science, and Technology,
Japan. KK, CHL, AI, HE, TS, HF and YK are also supported by (b) and HF and YK also acknowledge support from
MEXT, Global COE program of Chiba University. RWH, ASC and EMF acknowledge financial support from the
U.K. EPSRC. JSW acknowledges financial support from the Swiss National Centre of Competence in Research
program `MaNEP'. We are grateful for support and allocated beam time on D11 and D22 at the Institut Laue-
Langevin, Grenoble, France and on SANS-I at the spallation neutron source SINQ, Paul Scherrer Institut, Villigen,
Switzerland.


\begin{figure}[htpd]
\includegraphics[width=10cm]{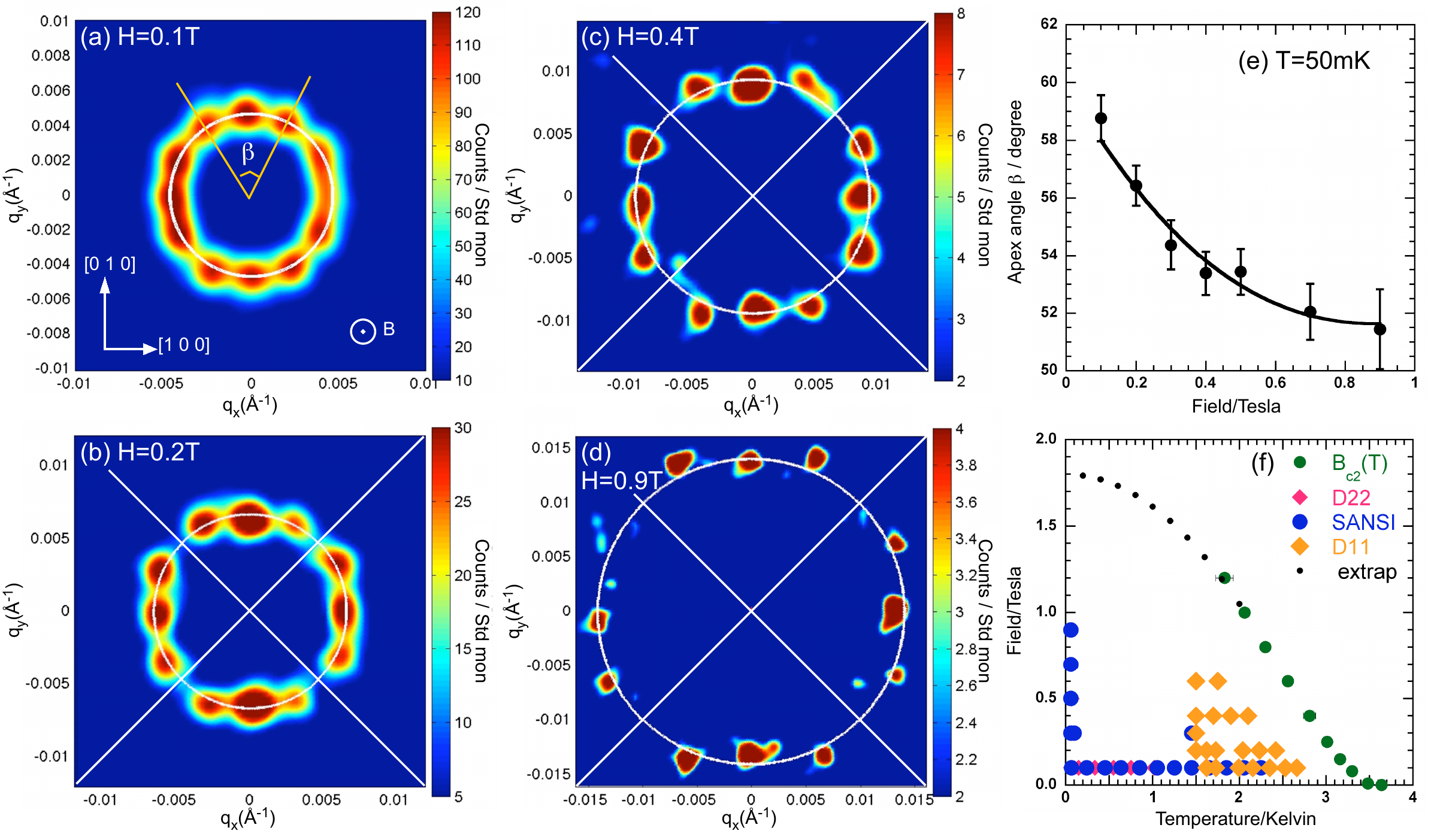}
\caption{\label{Fig1.eps}
Small angle neutron scattering patterns, VL structure and $B$-$T$ phase diagram for KFe$_2$As$_2$. a-d:
Diffraction patterns from the VL under the following conditions of applied field and temperature: (a) 0.1 T, 50 mK;
(b) 0.2 T, 1.5 K; (c) 0.4 T, 1.5 K; (d) 0.9 T, 50 mK, obtained on instruments D11 and D22 at ILL and SANS-I at PSI.
The diffracted intensity of a spot could be maximized by tilting the sample and field together onto the Bragg angle
~\cite{RS}. Backgrounds obtained without VL present were subtracted. To improve visibility,
the data were smoothed with a Gaussian of width comparable to the instrument resolution, and Poisson noise near
the main beam was masked. The higher field patterns are mosaics of results taken at the maximum intensities of
four sets of three spots, and at 0.1 T a sum over tilts is shown. White circles in the figures indicate the distance
from the main beam at which regular hexagonal VL spots would appear. The offsets of the spots from the circles
indicate that the VL is somewhat distorted from a pure hexagonal structure. (e): Field dependence of the apical
angle, $\beta$, of the VL structure at $T =$ 50 mK. (for definition of $\beta$ see Figure 1a). (f): The $B_{c2}$-$T$ phase diagram (from magnetization data with H $\parallel$ [0 0 1]) indicating all conditions under which VL measurements were
carried out. }
\end{figure}

\begin{figure}[htpd]
\includegraphics[width=6cm]{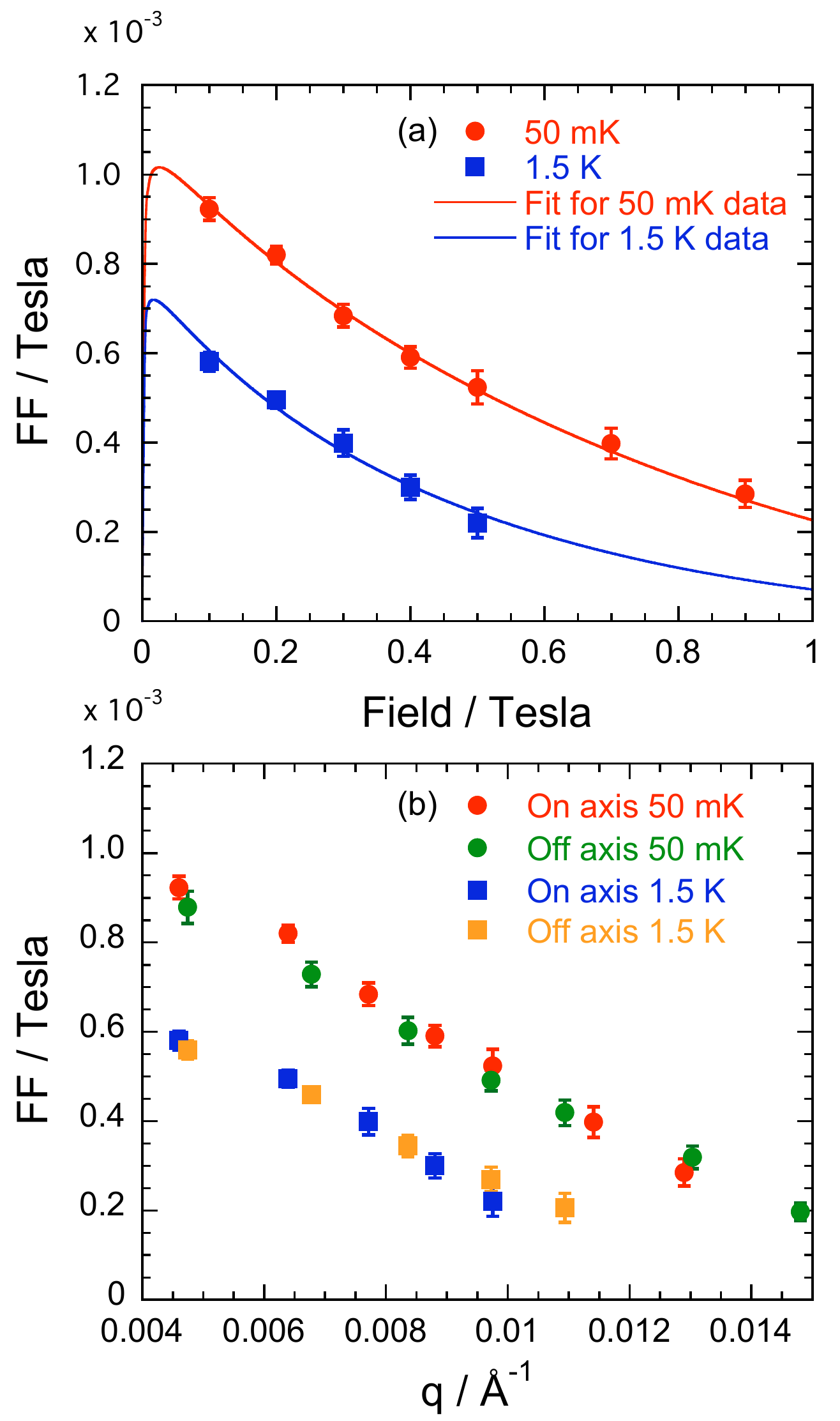}
\caption{\label{Fig2.eps}
(a): Field dependence of the Form Factor $(F)$ of the `on axis' VL diffraction spots at $T =$ 50 mK and 1.5 K. The
theoretical fit is described in the text. (b): $F$ for on- and off-axis spots at 50 mK and 1.5 K (plotted versus $q$-values, which are different for the two types of spot). }
\end{figure}

\begin{figure}[htpd]
\includegraphics[width=10cm]{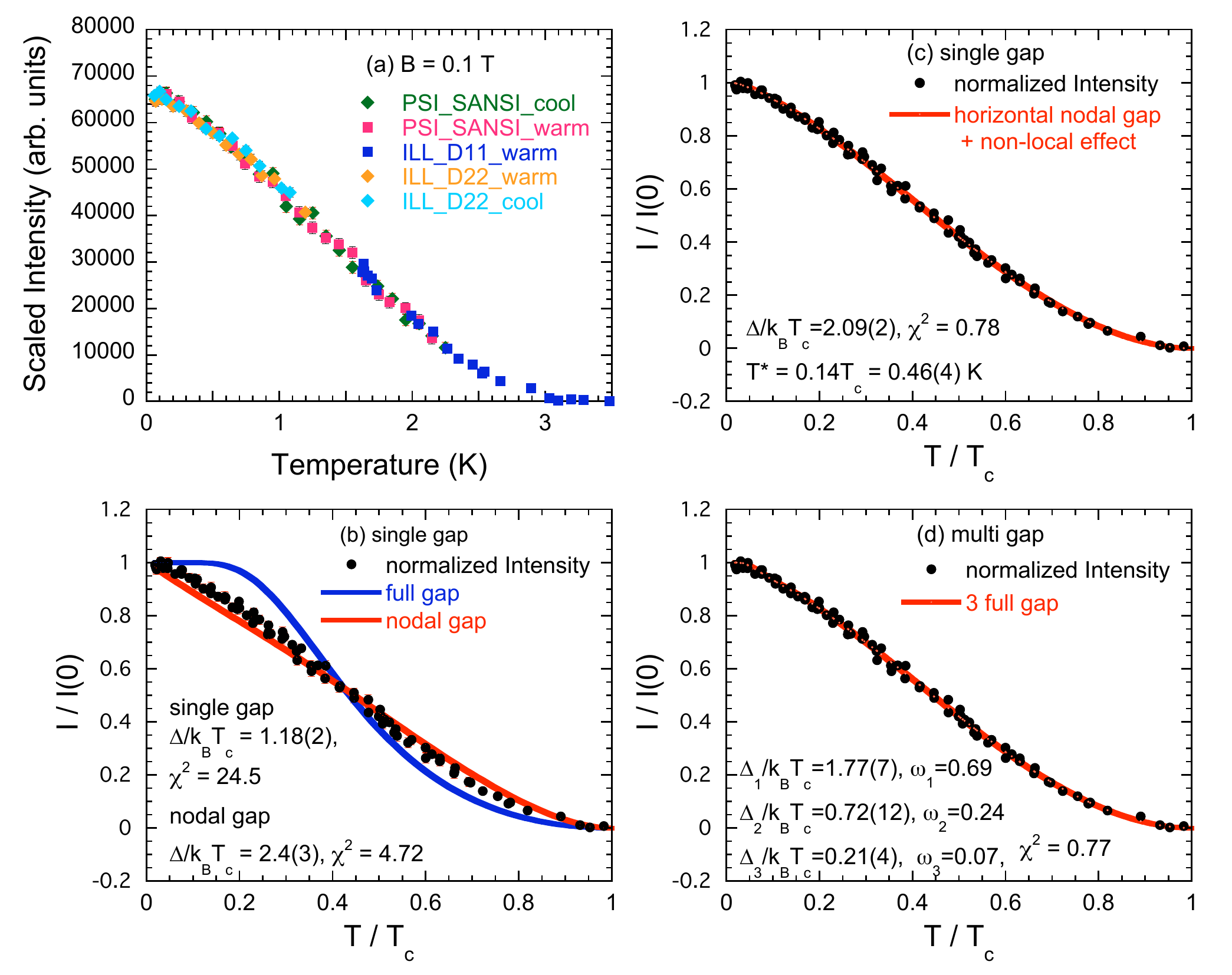}
\caption{\label{Fig3.eps}
(a): Temperature dependence of peak intensity at $B =$ 0.1 T. The sample tilt was fixed at a value maximizing the
diffracted intensity in the `top' spots, and data were taken versus either increasing or decreasing temperature. The
results from SANS-I were the most complete; the data taken at other instruments were included after scaling in
the overlapping regions~\cite{RS}. It is clear that all temperature dependencies are in
agreement within statistical errors in all cases, indicating that the system is in the thermal equilibrium state and the vortices are not relaxing to a new state on temperature cycling. (b-d): Fits of the data to various gap functions
versus $T / T_{c}$, with $T_{c}$ set to the value $T_{c2}$(0.1 T) $=$ 3.25 K. The fitted parameters are shown
inset in the graphs: (b) single full gap and single nodal gap; (c) single nodal gap including nonlocal effects~\cite{RS}. These cause a crossover from $T$-linear to $T^2$ behavior below a temperature $T^{\star}$~\cite{R23}. Similar effects can occur due to impurity scattering~\cite{R25A}, but are not expected to be strong in our clean material.}
\end{figure}

\end{document}